\documentclass{moriond}

\def\be{\begin{equation}}
\def\ee{\end{equation}}
\def\bea{\begin{eqnarray}}
\def\eea{\end{eqnarray}}

\newcommand{\Photo}{
\vspace*{0.5em}
\includegraphics[height=35mm]{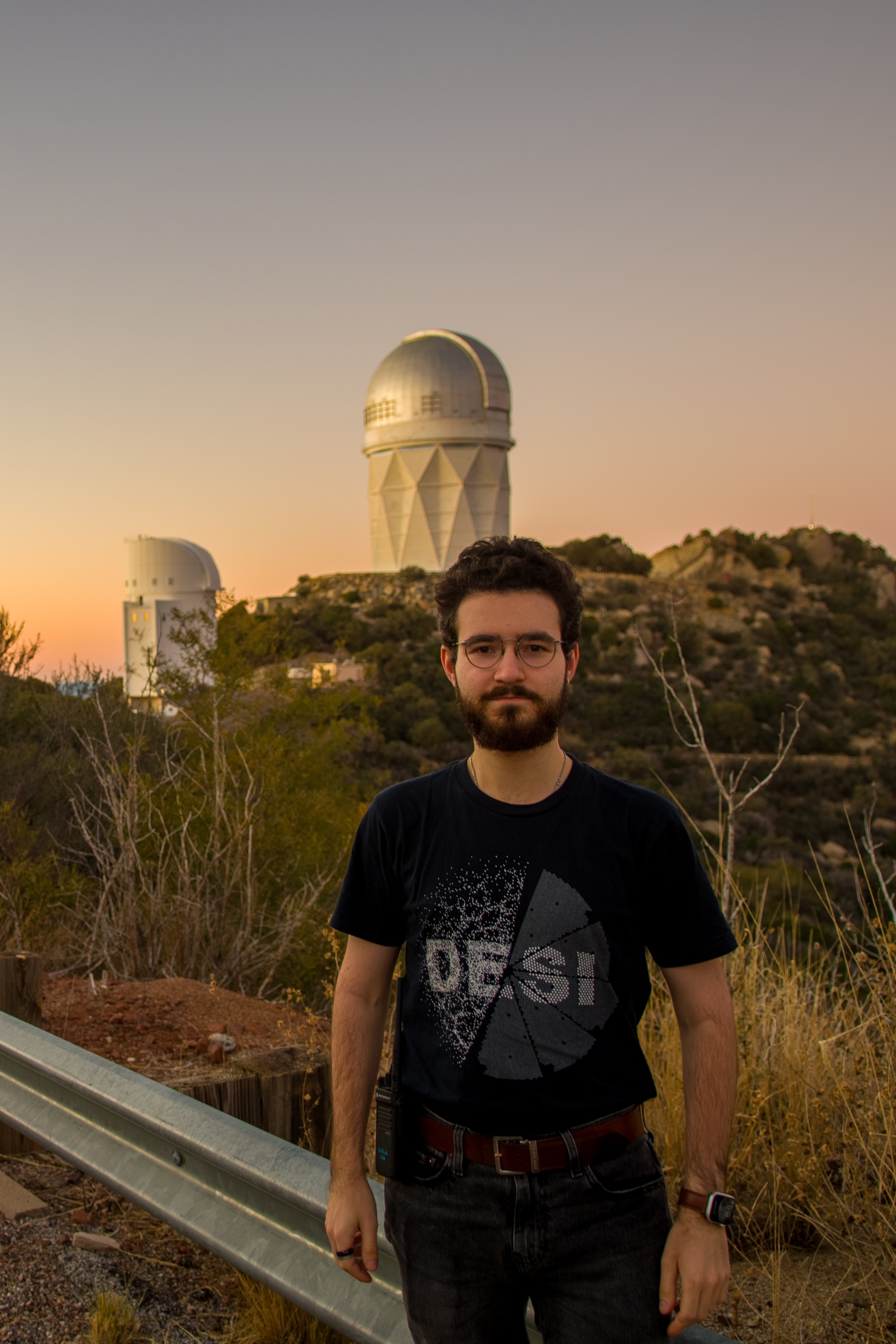}
\vspace*{-1em}
}

\usepackage{wrapfig}

\begin{document}
\vspace*{4cm}
\title{Optimal summary statistics to constrain $w_0 w_a$-CDM with the DESI BGS}

\author{ Simon Bouchard, Pauline Zarrouk }

\address{Laboratoire de Physique Nucléaire et des Hautes Energies, \\ Sorbonne Université}

\maketitle\abstracts{
The Dark Energy Spectroscopic Instrument (DESI) is the first stage-4 spectroscopic instrument installed on the 4-m Mayall Telescope at Kitt Peak National Observatory in Arizona. It contains 5000 robotically positioned fibers, allowing the measurement of as many galaxy spectra in one observation ($\sim$20 minutes). The data acquisition (2021--2028) aims at constraining the cosmological parameters, and especially the parameters describing the evolution of the dark energy with time.
However, standard clustering analysis based on 2-point statistics only capture the Gaussian information of the matter density field, while the analytical models used to match these statistical probe are limited to quasi-linear scales, as they are based on perturbation theory.
Therefore, the non-Gaussian and non-linear information of the density field is not exploited. 
To go beyond these limitations, several \textit{alternative} clustering methods have been developed that rely on a simulation-based pipeline.
In this presentation, we focus on the DESI Bright Galaxy Survey, a magnitude-limited sample of bright galaxies in the redshift range of $0.1 < z < 0.4$.
We present how accessing the non-linear and non-Gaussian information of the density field allows around 40\% improvement on the constraints of some $\Lambda$CDM parameters. We are also extending the method to constrain $w_0w_a$CDM in order to improve DESI's sensibility at low redshift where the Universe is dominated by dark energy.
}

\section{Introduction}
The BGS~\cite{Hahn_2023} is one of the densest available tracers in DESI and the closest to us, with a number density that is a factor 100 higher than the Luminous Red Galaxy (LRG) sample. 
Because the BGS probes a small volume of the Universe, the uncertainties on the cosmological parameters are dominated by a volume sample variance (or cosmic variance) and not by statistics. The current approach of the DESI Data Release 2 (DR2) standard analysis is to keep only 30\% of the BGS galaxies with an absolute magnitude cut $Mr<-21.35$~\cite{DESI_BAO_DR2}, as the sample variance does not allow improvements on the cosmological parameters. This limitation motivates the development of more innovative analysis methods that have the potential to exploit the full BGS statistics, including multi-tracer based on color cuts~\cite{Trusov_thesis_2024} or on local density. The latter is referred to as density-split clustering and is one of the alternative clustering methods beyond 2-point statistics that we explore in this presentation.

\begingroup
\setlength{\columnsep}{1em}%

\begin{wrapfigure}{R}{0.45\textwidth}
    \centering
    \includegraphics[width=1.0\linewidth]{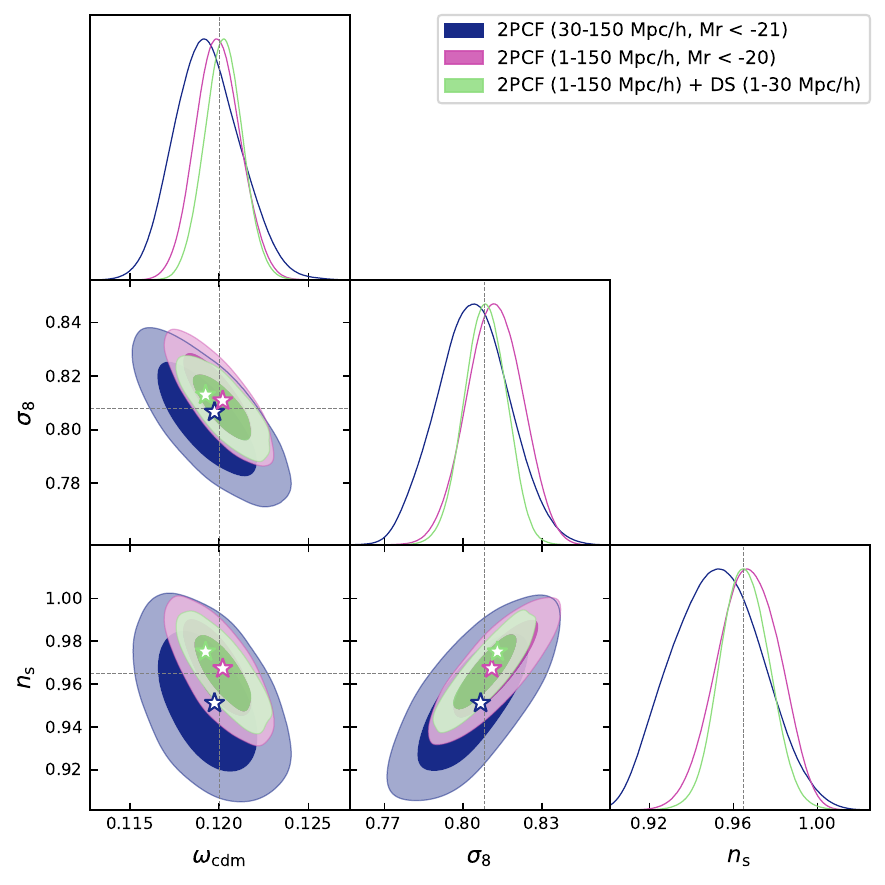}
    \caption{
    Constraint on the Cosmological parameters (resp. $\omega_{cdm}$, $\sigma_8$ and $n_s$), marginalized over the astrophysical parameters. The blue contour shows the DESI DR-1 baseline with only the 2PCF constraint and a magnitude cut at $Mr < -21.35$; the red contour shows the constraint of the 2PCF with the small scales accessed trough the emulator with a magnitude cut at $Mr < -20$; the green contour shows the final constraint, adding the density-split model.
    }%
    \label{fig:validation}
\end{wrapfigure}

\section{Alternative Clustering Methods}
The \textit{Alternative Clustering Methods} topical group in DESI developed several methods to access the non-Gaussian information. We focus here on the Density-Split method~\cite{Paillas_2021}, which computes a local over-density histogram for each tracer in a galaxy catalog, and splits the catalog in a fixed number of quantiles. By computing the cross-correlation of each quantiles with the rest of the catalog, we access the non-Gaussian information of the matter density field~\cite{Paillas_2023}. This approach also can reduce the sample volume variance that limits the precision on the cosmological parameters\cite{Abramo_2013} and has already been applied to BOSS LRG~\cite{Paillas_2024}.
This pipeline uses a simulation-based model that relies on neural networks (using \texttt{sunbird}~\cite{Cuesta_Lazaro_2023}) to reproduce the statistics at small scales, allowing the access to non-linear information. A training set of galaxy catalogs is generated using the \texttt{AbacusSummit}~\cite{Maksimova_2021} suite of Dark Matter simulations, and emulators are trained to learn the dependence of different statistics with the cosmological and astrophysical (HOD~\cite{Yuan_2021}) parameters.
Applying this pipeline to the BGS sample raises several computational challenges, as the constraints on the computational time and storage space scale with the density of the sample.

\section{Results}
By choosing a new BGS magnitude cut at $Mr<-20$, we recover $\sim$70\% of the data points in the BGS sample. The figure~\ref{fig:validation} shows the constrains on the cosmological parameters on validation mocks when using this new BGS definition, and adding the density-split constraint compared to the DR-1 BGS analysis baseline. We recover cosmological parameters within 1$\sigma$, with up to 40\% improvement on the intrinsic model constraint. Similar results are obtained when testing this inference pipeline on DESI cubic mocks~\cite{Smith_2024} reproducing the BGS sample. Extending the method to the $w_0 w_a$-CDM model shows even better constraints, but the AbacusSummit prior on the cosmological parameters priors limits the parameter space. This limitation motivates the development of new cosmological simulation suites to better exploit the information contained in the available data.

\endgroup

\section*{References}
{\footnotesize
\bibliography{moriond}

\begin{thebibliography}{10}

\bibitem{Hahn_2023}
ChangHoon Hahn et~al.
\newblock {\em The Astronomical Journal}, 165(6):253, May 2023.

\bibitem{DESI_BAO_DR2}
DESI Collaboration et~al.
\newblock {\em Physical Review D}, 112(8), October 2025.

\bibitem{Trusov_thesis_2024}
Svyatoslav Trusov.
\newblock Theses, {Sorbonne Universit{\'e}}, September 2024.

\bibitem{Paillas_2021}
Enrique Paillas et~al.
\newblock {\em MNRAS}, 505(4):5731–5752, June 2021.

\bibitem{Paillas_2023}
Enrique Paillas et~al.
\newblock {\em MNRAS}, 522(1):606–625, April 2023.

\bibitem{Abramo_2013}
L.~Raul {Abramo} et~al.
\newblock {\em MNRAS}, 432(1):318--326, June 2013.

\bibitem{Paillas_2024}
Enrique Paillas et~al.
\newblock {\em MNRAS}, 531(1):898--918, 06 2024.

\bibitem{Cuesta_Lazaro_2023}
Carolina {Cuesta-Lazaro} et~al.
\newblock {\em MNRAS}, 531(3):3336--3356, July 2024.

\bibitem{Maksimova_2021}
Nina~A Maksimova et~al.
\newblock {\em MNRAS}, 508(3):4017–4037, 2021.

\bibitem{Yuan_2021}
Sihan Yuan et~al.
\newblock {\em MNRAS}, 510(3):3301–3320, November 2021.

\bibitem{Smith_2024}
A~Smith et~al.
\newblock {\em MNRAS}, 532(1):903–919, 2024.

\end{thebibliography}
}

\end{document}